\documentclass[12pt]{article}
\usepackage{amssymb,amsmath}
\usepackage{graphicx}  
\begin{document}
\title{Antiprotonic hydrogen in static electric field}
\author{G.Ya. Korenman and S.N. Yudin\\[1mm]
\emph{Skobeltsyn Institute of Nuclear Physics, Moscow State 
University,} 
\\ \emph{Moscow, Russia}  
\\ e-mail: korenman@anna19.sinp.msu.ru} 
 \date{ }
\maketitle 
\begin{abstract}
 \noindent
Effects of the static electric field on the splitting and annihilation 
widths of the levels of antiprotonic hydrogen with a large principal 
quantum number ($n=30$) are studied. Non-trivial aspects of the 
consideration is related with instability of $(p\bar{p})^*$-atom in 
$ns$ and $np$-states due to coupling of these states with the 
annihilation channels.  Properties of the mixed $nl$-levels are 
investigated depending on the value of external static electric field. 
Specific resonance-like dependence of effective annihilation widths on 
the strength of the field is revealed. 
\end{abstract}

\section{Introduction}
When antiproton enters a matter it is slowed down and captured 
eventually by atom forming an antiprotonic atom. From the general point 
of view this new physical object is of the great interest. For the long 
time the experimental methods to investigate antiprotonic atoms have 
been limited by studying their X-ray radiation and the products of 
antiproton annihilation at the end of cascade transitions. However, 
about 15 years ago the investigations of these atoms have got on the 
new qualitative level especially due to the new experimental methods 
(see \cite{1,2} and references therein) that used antiprotonic beam 
from storing ring of  LEAR operated in CERN up to 1996. New setup AD 
("antiproton decelerator") that has been operating since the end of 
2001 opens, in definite aspects, even more possibilities  for 
experimental study of antiprotonic atoms. In particular, in the frame 
of ASACUSA project \cite{3} it is supposed, among other experiments, 
direct observation and investigation of antiprotonic atoms formation in 
very rarefied gases. 

The new experimental possibilities pose various problems of the theory 
of antiprotonic atoms that were not considered early because they 
seemed to be of little importance in the former experimental 
conditions. One of such problem could be an influence of external 
static electric field on the properties of antiprotonic atoms, 
especially antiprotonic hydrogen. Static electric and magnetic fields  
can be generated by some parts of an experimental setup, therefore they 
have to be taken into account in precision measurements of the 
properties of excited $(p\bar{p})$ atom. On other hand, a special 
choice of the fields could be used in order to obtain an additional 
information on the properties of antiprotonic atoms.  

According to current concepts of the theory of exotic atom formation 
(see, e.g., \cite{4,5}), the antiprotonic atoms are formed initially in 
highly exited states with principal quantum number $n\geqslant  n_0 
\simeq \sqrt{\mu}$ and, predominantly, with large values of orbital 
angular momenta ($1\ll l\leqslant n-1$), where $\mu$ is a reduced mass 
of the $\bar{p}$ - nucleus system in the units of electron mass (so, 
for the formation in atomic hydrogen $\mu=918,\, n_0=30$). When 
$(p\bar{p})_{nl}$ atom is formed in molecular hydrogen, the 
distribution over quantum number is more complicated, however the 
probability of population of the states with large $n$ and $l$ remains 
rather appreciable. Among the highly excited states, specific interest 
can present circular orbits with $l=n-1$ and nearly-circular orbits. In 
isolated antiprotonic hydrogen atom these states can decay only by 
radiative E1-transition. Radiative life time of hydrogen-like atom at 
large $n, l$ can be estimated by the equation \cite{6}
\begin{equation} \label{eq1}
 \tau_{\gamma}(nl) \simeq (n^3 l^2/\mu Z^4)\cdot 84.5 \,\mathrm{ps}
\end{equation}
that gives $\tau_\gamma = 2.09\,\mu$s  for the circular orbit with 
$n=30$ of antiprotonic hydrogen. This life time is of the same order of 
value as for antiprotonic helium metastable states \cite{1}, therefore 
the same high-precision method of laser-induced transition could be, in 
principle, applied for the study of energy  structure and dynamics of 
the excited states of antiprotonic hydrogen. However,  the $(p\bar{p})$ 
atom, as compare with $(\bar{p}\mathrm{He}^+)$,  may be less stable 
against atomic collisions and influence of external fields, because a 
most part of the  $(p\bar{p})_{nl}$ states with a definite $n$ is 
degenerated in $l$, contrary to antiprotonic helium. Therefore 
theoretical study of the quenching effects on the long-lived states of 
antiprotonic hydrogen is important for the possible future experiments. 
One of the most important factors to quench long-lived states is the 
electric field arising in the processes of atomic collisions or as a 
results of the external conditions. The dynamical (collisional) Stark 
effect in hadronic hydrogen atoms was discussed in the literature for a 
long time \cite{6,7}. However the influence of the external static 
electrical field, as far as we know, was not discussed in the 
literature, in spite of the seeming simplicity of this effect. In this 
paper we discuss the influence of electric field on the splitting and 
decay rates of highly-excited ($n=30$) states of antiprotonic hydrogen 
atoms. 

\section{Formulation of the problem}
The states with large quantum numbers $n,l$ of isolated antiprotonic 
hydrogen $(p\bar{p})_{nl}$ are long-lived, because annihilation in the 
states with $l\gg 1$ is practically absent and, on other hand, these 
states have large radiative life times. However, if some external 
action forces transitions of the  $(p\bar{p})$ system to the states 
with low orbital angular momenta, then antiproton will annihilate 
quickly. In the problem under consideration an external action is 
produced by the static external electric field. Formally this problem 
is formulated as follows.

If the nuclear $(p\bar{p})$-interaction is disregarded, all the states 
$nl$ of antiprotonic hydrogen atom will be degenerated in $l$. In the 
real system nuclear interaction is important in the states with low 
angular momenta and it produces the both shifts and annihilation widths 
of the  levels, or, in other words, complex shifts
 \begin{equation} \label{eq2}
\Delta E_{nl}=- \epsilon_{nl} - \mathrm{i}\Gamma_{nl}/2 . 
 \end{equation}
The estimations show that these values at $n\sim 30$ are important for  
$l=0,\, 1$ and can be neglected for $l\geq 2$. So, the complex shifts 
remove the degeneration of the $ns$ and $np-$levels, whereas other 
levels remain to be degenerated. A dependence of the complex shifts on 
the principal quantum number can be obtained taking into account that 
the radius of annihilation region and of nuclear interaction is small 
as compare with a radius of the antiprotonic hydrogen atom. Radial wave 
function of antiprotonic atom this region should have the simple form 
$R_{nl}(r)\simeq C_{nl}\cdot r^l$. It allows to express the dependence 
the complex shifts $\Delta E_{nl}$ on $n$ in terms of the coefficients 
$C_{nl}$: 
\begin{equation} \label{eq3}
\Delta E_{nl} = \Delta E_{l+1,l} |C_{nl}/C_{l+1,l}|^2 
\end{equation}

The coefficients $C_{nl}$ are estimated, as a rule, with the 
hydrogen-like wave functions that gives
\begin{align}
\Gamma_{ns} & = \Gamma_{1s}/n^3, \label{eq4} \\
 \Gamma_{np}& =\Gamma_{2p}\cdot 32(n^2-1)/(3 n^5), \label{eq5} 
\end{align}
and similarly for $\epsilon_{nl}$. As the input data we have taken the 
following values for the energy shifts and widths \cite{8}: 
\begin{equation} \label{eq6}
\Gamma_{1s}/2 =561 \text{ eV}, \quad \epsilon_{1s}=691 \text{ eV} , 
\qquad \Gamma_{2p}/2= 17 \text{ meV}, \quad \epsilon_{2p}=0. 
\end{equation}
With these values we obtain from the Eqs. \eqref{eq4}, \eqref{eq5} for 
the states with $n=30$: 
\begin{equation} \label{eq7}
\Gamma_{ns}/2 =20.78 \text{ meV}, \quad \epsilon_{ns}=25.59 \text{ 
meV}, \qquad \Gamma_{np}/2 =0.00671 \text{ meV},  \quad 
\epsilon_{np}=0. 
\end{equation}

Total hamiltonian $H$ of the antiprotonic hydrogen atom in the external 
electric field can be written as a sum of two terms, 
\begin{equation} \label{eq8}
    H=H_0 + V,
\end{equation}
where $H_0$ is a diagonal matrix with elements 
\begin{equation}  \label{eq9}  
\langle nl|H_0|nl'\rangle  = \delta_{ll'}\left( \delta_{l0} 
\Delta{E_{ns}} + \delta_{l1} \Delta{E_{np}} \right) . 
\end{equation}
The matrix of interaction of the atom with the electric field $E$ 
directed along $Oz$ axis has the form: 
\begin{equation} \label{eq10}
\langle nlm|V|nl'm' \rangle = eE r_0 (3/2) n \sqrt{n^2-l_>^2} 
 \langle l010|l'0 \rangle \langle l'm'10|lm\rangle ,
\end{equation}
where $l_>=\max(l,l')$, the quantities like $\langle l010|l'0 \rangle$ 
are Clebsh-Gordan coefficients, $r_0= r_B /\mu$,  $\mu=M_p/2$ is a 
reduced mass of the $(p-\bar{p})$ system in the units of electron mass, 
and $r_B$ is the Bohr radius of ordinary hydrogen atom. 

Eigenvalues of the non-hermitian matrix \eqref{eq8} give the complex 
energies of the $(p\bar{p})_n$ states in the static electric field. The 
real parts of these energies are shifts of the levels from the 
unperturbed values and the imaginary parts are the annihilation widths 
with account for a mixing of the states with different $l$ due to 
electric field. 

\section{Results and Discussion}
The formulation of the problem, as given in the previous section, 
supposes that a mixing of the states with different $l$ at fixed $n$ is 
important, whereas a mixing of the states with different $n$ can be 
neglected. It means that our consideration has to be restricted by the 
field less than a critical value, at which non-diagonal matrix element 
$\langle nl|V|n'l'\rangle$ is comparable with the distance between the 
levels with different $n$. It is easy to estimate that for the levels 
with $n\simeq 30$ this critical value of the field is $E_a\sim 10^9$ 
V/cm, that is far from real laboratory fields. 

It should be noted that in the problem under consideration there are 
also two other critical values of the field. The first one is the field 
that mixes effectively $ns$ and $np$ states, and other one is that 
mixed $np$ and $nd$ states, the latter being degenerated with all other 
$nl$ states for $l>2$. With the above-mentioned values of strong 
interaction shifts and annihilation widths of the $ns$ and $np$ states, 
we estimate the corresponding critical fields as $E_b \sim 3.3\cdot 
10^6$ V/cm  for a mixing of $ns$ and $np$ at $n=30$, and $E_c\sim 2000$ 
V/cm for a mixing of $np$ and $nd$ states. We restrict our 
consideration by the fields up to 5000 V/cm, therefore only third 
critical value $E_c$ of the field is covered by these calculations. 

For the further discussion of the obtained results let us remember 
also, that the problem of Stark mixing in the degenerated (without 
initial shifts and widths) hydrogen-like system at the fixed $n$ has 
the exact solution in the parabolic coordinates \cite{9}. Eigenstates 
of a such system  are labeled by parabolic quantum numbers $n_1,n_2,m$ 
($n_1+n_2+|m|+1=n$), the splitting of the levels is linear in electric 
field, whereas the coefficients $\langle nlm|n_1 n_2 m \rangle$ of 
mixing of the states with different $l$ do not depend on the value of 
the field and can be expressed in terms of Clebsh-Gordan coefficients,  
\begin{equation} \label{eq11}
\langle nlm|n_1 n_2 m \rangle^2= \langle k\nu_{1}k\nu_{2}|lm \rangle^2, 
\end{equation}
where $k=(n-1)/2$, $\nu_1=(n_1- n_2 +m)/2$, $\nu_2=m-\nu_1$. Therefore 
the admixture of $ns$-state to all parabolic states with $m=0$ should 
be $\langle ns|nn_1 \rangle^2 =1/n$, and the weights of the np-state in 
the parabolic states are $\langle np,m|nn_1,m\rangle ^2 = 0.019 \div 
0.025$ at $n=30$. These relationships could describe the effect of 
external field on antiprotonic hydrogen at very high values of the 
electric field ($E_b< E <E_a$). However for the realistic values of $E< 
E_b$ the influence of the field can not be predicted without 
calculations. 

\begin{figure}[thb]
\begin{center}
\includegraphics[width=7.75cm, height=5.5cm]{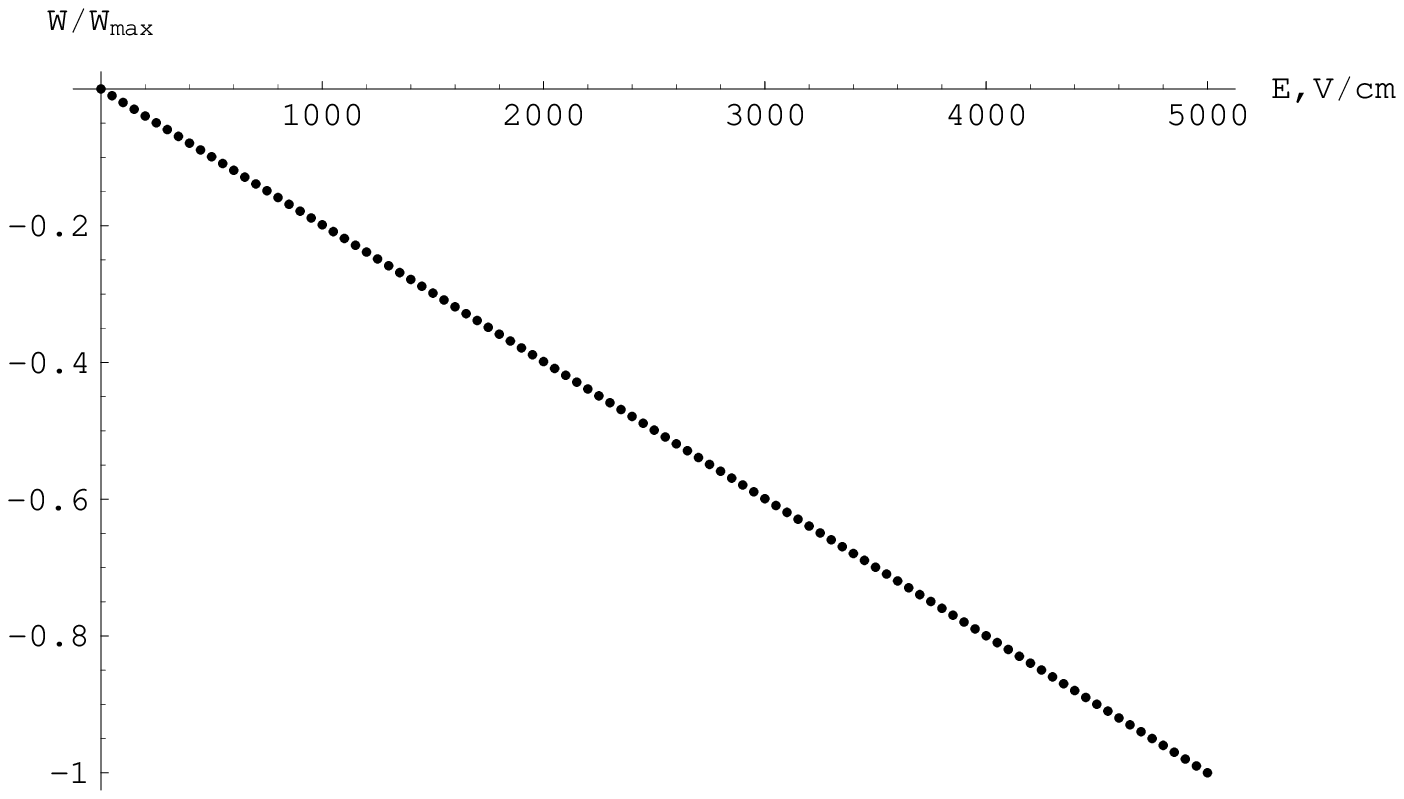}
\put(-100,105){(a)} 
\includegraphics[width=7.75cm, height=5.5cm]{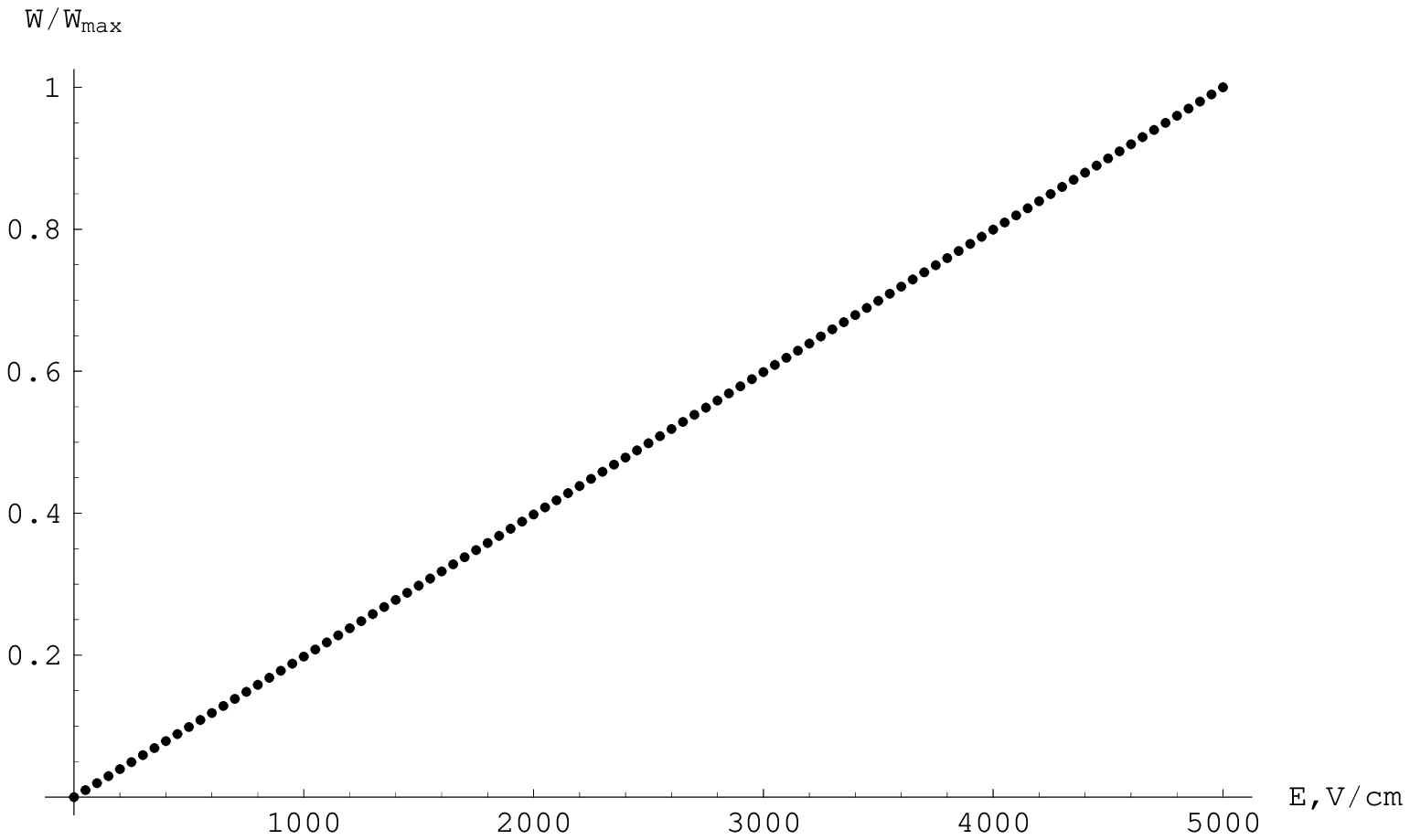}
 \put(-100,105){(b)}
 \caption{The dependence of the shifts of $(p\bar{p})$ states on 
electric field.  Part (a): $l=1$, $W_{\max}= 3.7\cdot 10^{-5}$ eV; part 
(b): $l=28$, $W_{\max} =3.43\cdot 10^{-5}$ eV.} 
 \label{fig1} 
\end{center}
\end{figure}

\begin{figure}[thb]
\begin{center}
\includegraphics[width=7.75cm, height=5.5cm]{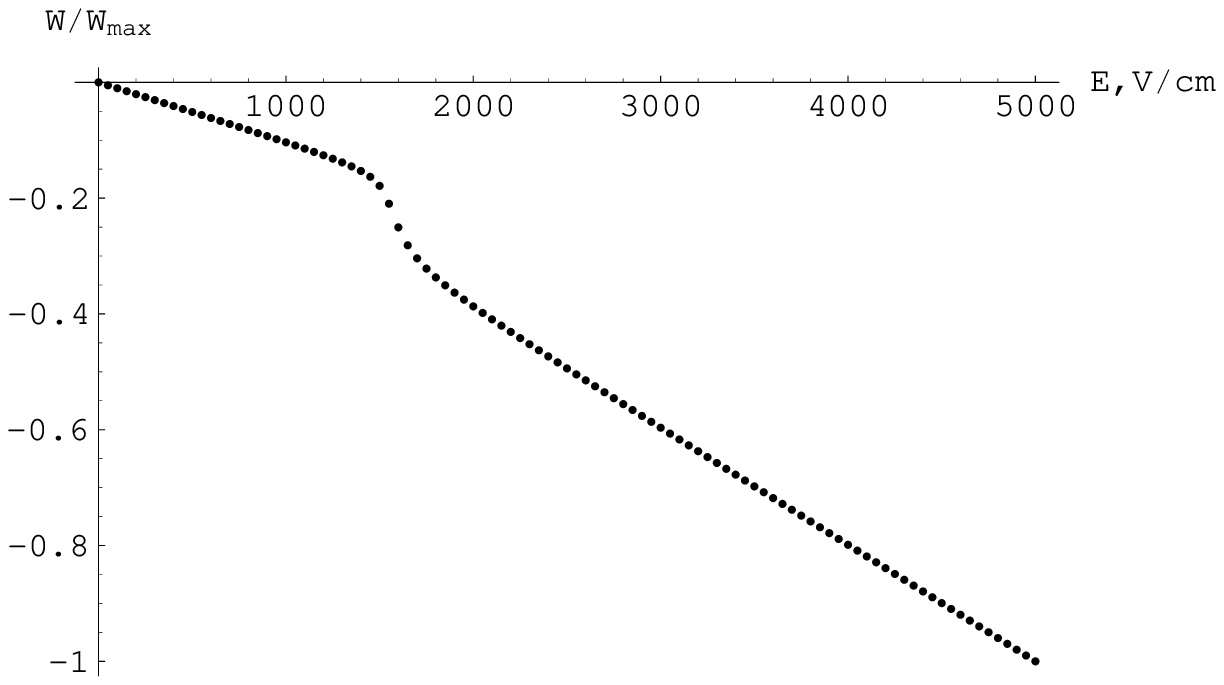}
\put(-100,105) {(a)} 
\includegraphics[width=7.75cm, height=5.5cm]{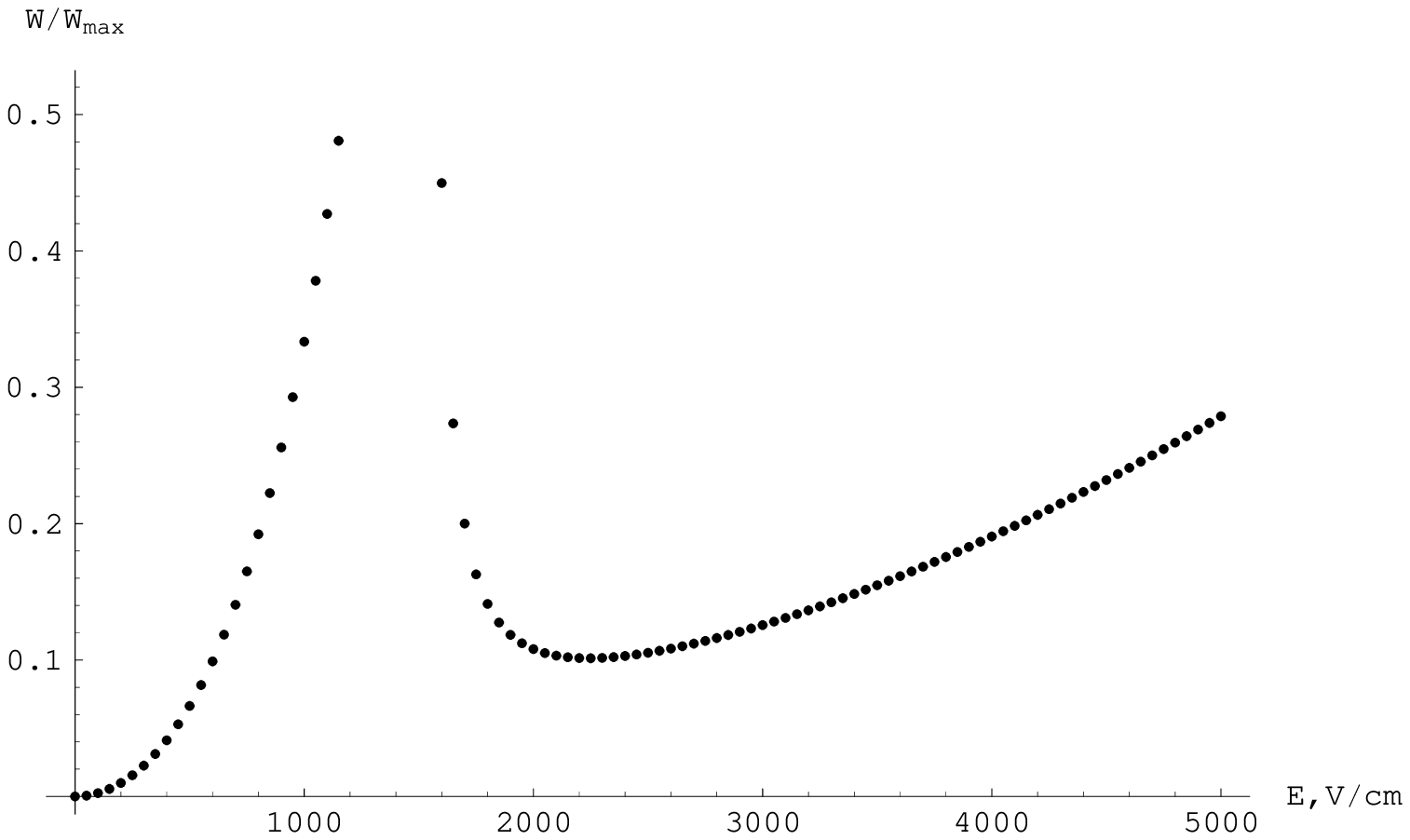}
 \put(-100,105) {(b)} 
\caption{Resonance-like dependence of the shifts of $(p\bar{p})$ states 
on electric field. Part (a): $l=14$, $W_{\max} =2.62\cdot 10^{-6}$ eV; 
part (b): $l=15$,  $W_{\max}=1.95\cdot 10^{-9}$ eV} 
 \label{fig2} 
\end{center}
\end{figure}

\begin{figure}[thb]
\begin{center}
\includegraphics[width=7.75cm,height=5.5cm]{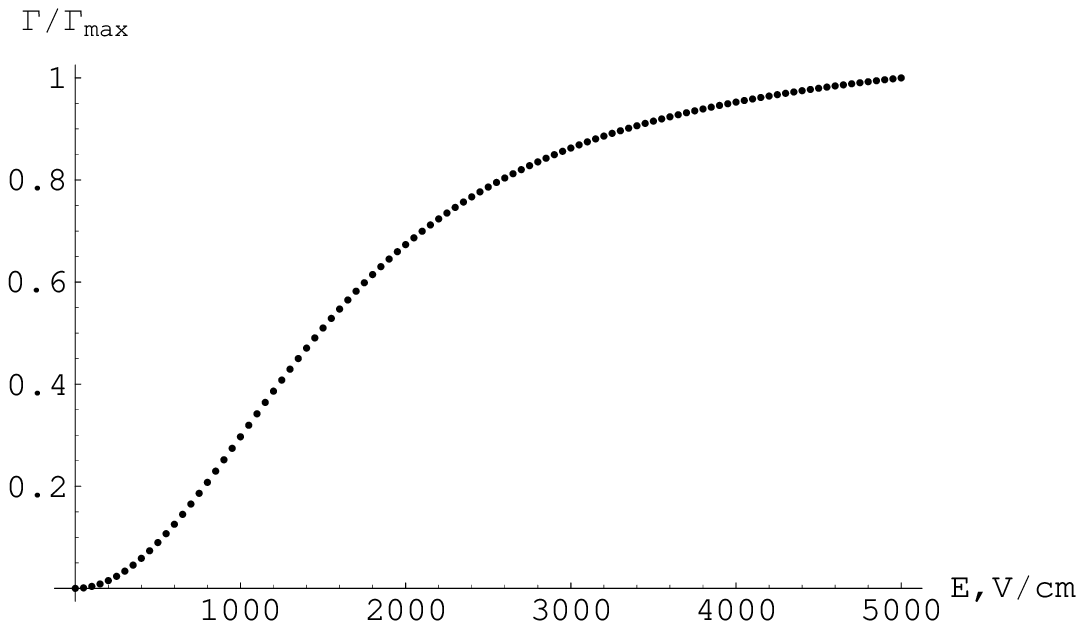}
\put(-100,145){(a)} 
\includegraphics[width=7.75cm,height=5.5cm]{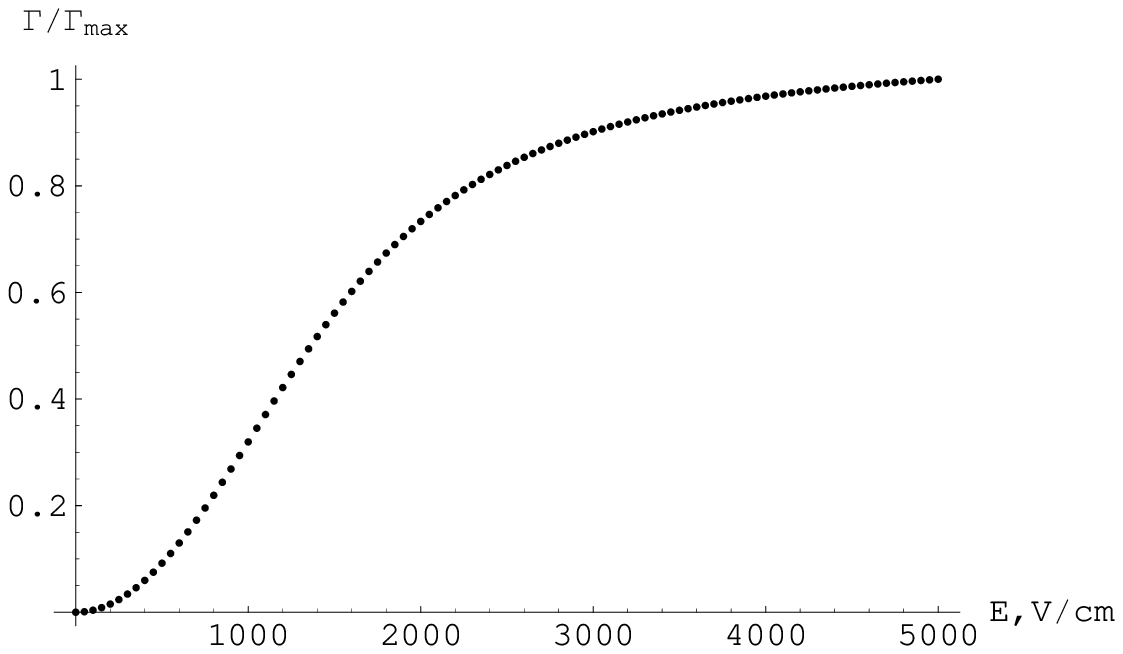}
 \put(-100,145) {(b)}
\caption{Dependence of the widths on the value of electric field. Part 
(a): $l=1$; $\Gamma_{\max} =3.42\cdot 10^8$ 1/s; part (b): $l=28$, 
$\Gamma_{\max} =4.52\cdot 10^8$ 1/s.} 
 \label{fig3} 
\end{center}
\end{figure}

\begin{figure}[thb]
\begin{center}
\includegraphics[width=7.75cm, height=5.5cm]{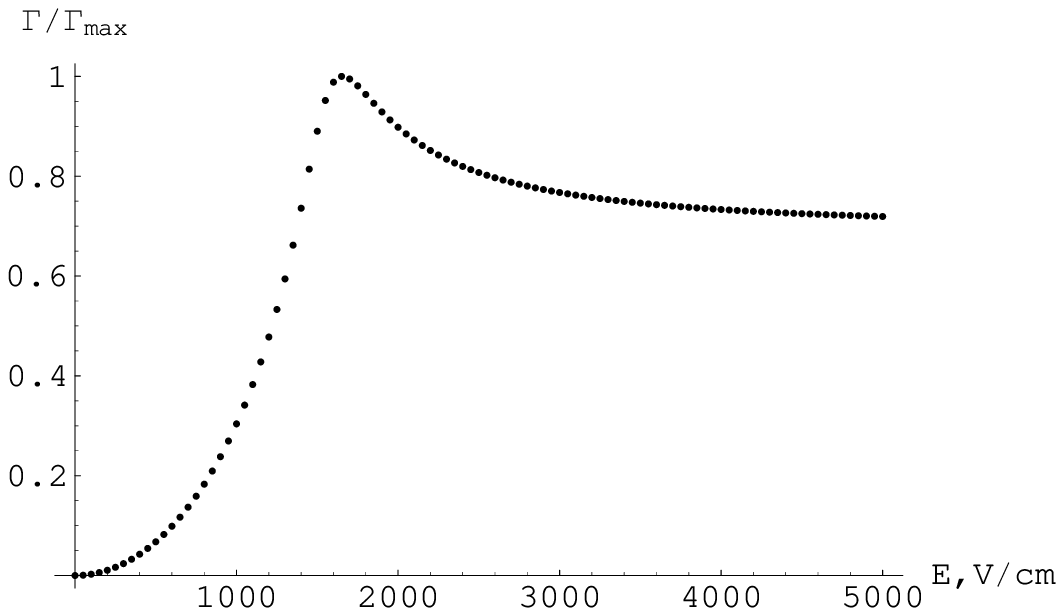}
\put(-100,145){(a)} 
\includegraphics[width=7.75cm, height=5.5cm]{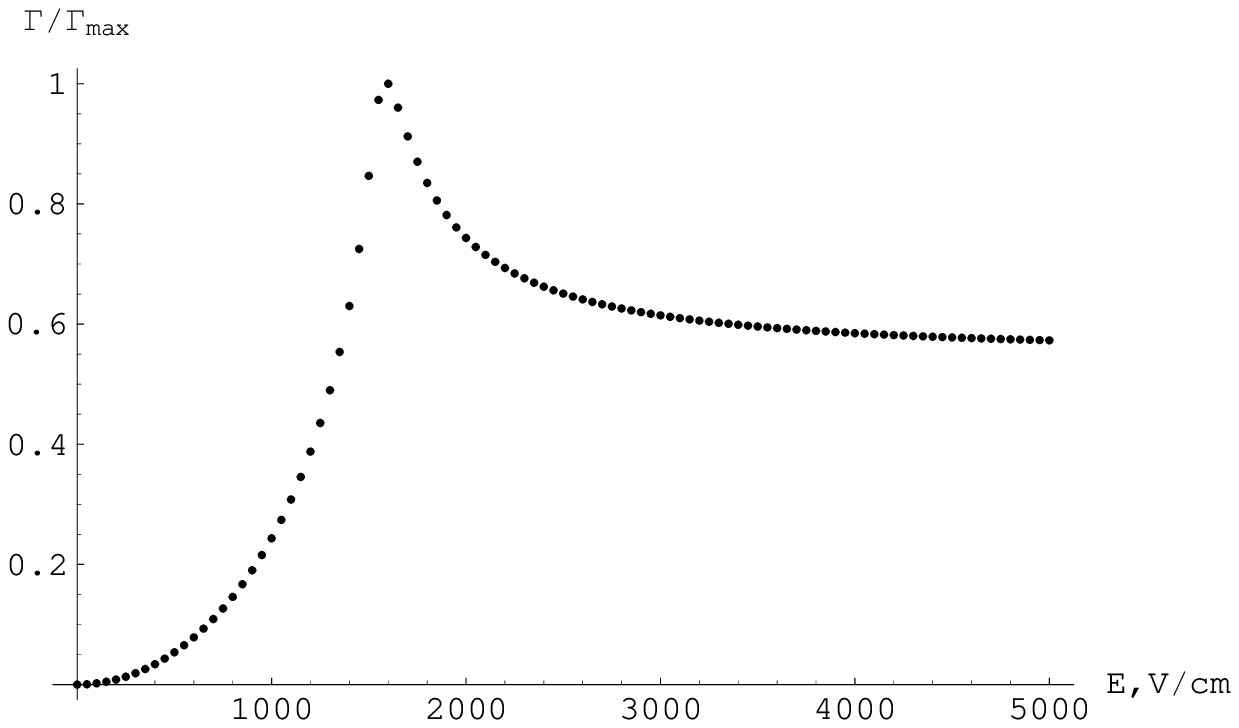}
 \put(-100,145) {(b)}\\ \caption{Resonance-like 
dependence of the widths on the value of electric field. Part (a):  
$l=14$, $\Gamma_{\max}=1.51\cdot 10^9$ 1/s; part (b): $l=16$, 
$\Gamma_{\max} =1.52\cdot 10^9$  1/s.} 
 \label{fig4} 
\end{center}
\end{figure}

The results of our calculations are shown on Figs. \ref{fig1} - 
\ref{fig4}. On these figures the shifts $W(E)$ and widths $\Gamma(E)$ 
of the states are presented in the relative units of $|W_{\max}|$ and 
$\Gamma_{\max}$, which are the maximum values of these quantities in 
the considered interval of electric fields ($0 \leq E\leq 5000$ V/cm). 

Different types of dependencies of the shifts on the value $E$ of 
electric field are shown on Figs. \ref{fig1} - \ref{fig2}  for several
levels with $m =0$. The levels are labeled by the indexes $l$ that 
correspond to the quantum numbers of the states in the limit $E=0$. 
With respect to the dependence of the shifts on $E$, the levels  can be 
divided into two main types. The shifts are negative for $l\leq 14$ and 
positive for $l\geq 16$, and in the both cases they rise in absolute 
value with increasing of the electric field. Within the intermediate 
interval $l\in 13-16$ the shifts are very small and change the signs. 
It is seen from Fig. \ref{fig2} that in this region the shifts are 
nonlinear in $E$ and, moreover, they behave themselves for the first 
sight unusually, showing a resonance-like behaviour at $E\sim E_c$. 
However, the nature of a such behaviour is quite clear - this is a 
result of some crossing of the levels that takes place with a change of 
the values of electric field. For the levels with $|m|=1$ we observe a 
similar behaviour with $E$. 

The widths of the levels with $m =0$ are shown on Figs. \ref{fig3} - 
\ref{fig4} depending on the value of electric field.  For the states 
with $l\leq13$ and $l\geq16$ the widths grow with increasing of the 
electric field, however in the aforementioned interval $13\geq l \leq 
16$ the widths depend on the field in a resonance-like manner. The 
reason of this phenomenon is the same as for the shifts, i.e. it is 
connected with the crossing of levels in the static electric field. 

The shift and width of the $ns$ state are practically unchanged at the 
considered values of $E$. It is related with the large values of the 
'self' shift and width, or, in other words, with the fact that the 
value of $E$ is small as compare with critical field $E_b$ as estimated 
above. Thus the obtained values of 'induced' widths of all the states 
with $l\neq 0$ are due to the influence of np-states in the considered 
interval of electric fields. At larger electric fields the contribution 
of $ns$ states would be a more essential. 

\section{Conclusion}

In this paper we have made the analysis of the splitting and 
annihilation widths of the levels of antiprotonic hydrogen atom in the 
static external electric field. As compare with the theoretical case 
when all the levels are degenerated and the exact analytical solution 
of this problem exists, we meet here a more complicated situation. This 
complication is related with the strong interaction shifts and 
annihilation widths of $ns$ and $np$ states participated in the mixing 
of $nl$ levels of the antiprotonic atom. The results of the paper can 
be formulated as follows: 

a) due to mixing of different $nl$ states ($l>1$) with $np$ states at 
the value of electric field $E\sim 1000$ V/cm, all the states acquire 
the widths of order $10^8\div 10^9$ 1/s. The relevant shifts for the 
most part of the states are linear in $E$ and have the order of value 
$10^{-5}$ eV. 

b) Contribution of the $ns$ state into the widths of other levels is 
negligible at the considered values of electric field ($E\leq 5000$ 
V/cm), because the $ns$ state is very distant in complex energy from 
all other levels and can be admixed to these states at $E\sim 3\cdot 
10^6$ V/cm. As a result, the main contribution to the widths of the 
mixed levels gives $np$ state. 

c) For the states that are close to the point, where the sign of the 
shift is changed, the shifts and widths of exact states have a 
resonance-like dependence on the value of electric field. The 
'resonance' value of electric field is of the same order as critical 
value $E_c$, which provides strong mixing of the $np$ state with other 
states ($l\geq 2$).  This phenomenon is due to a crossing of the levels 
at some value of the electric field. 

\section*{Acknowledgements}

This investigation was supported by Russian Foundation for Basic 
Research as a part of the project 03-02-16616. Authors thank to N.P. 
Yudin for the useful discussions. One of the authors (G.K.) thanks to 
Y. Yamazaki for attracting our interest to the considered problem.

\end{document}